\documentclass[aps,prl,twocolumn,showpacs,superscriptaddress,nofootinbib]{revtex4-1} 
\usepackage{graphicx}
\usepackage{capt-of}
\usepackage{amsmath}

\usepackage{amssymb}
\usepackage{relsize}

\usepackage{xcolor}

\usepackage{multirow}

\newcommand{\bra}{\langle}
\newcommand{\ket}{\rangle}
\newcommand{\bs}[1]{\ensuremath{\boldsymbol{#1}}}
\newcommand{\nn}{\nonumber\\}

\newcommand{\be}{\begin{equation}}
\newcommand{\ee}{\end{equation}}
\newcommand{\bea}{\begin{eqnarray}}
\newcommand{\eea}{\end{eqnarray}}

\DeclareMathOperator*{\SumInt}{%
\mathchoice%
  {\ooalign{$\displaystyle\sum$\cr\hidewidth$\displaystyle\int$\hidewidth\cr}}
  {\ooalign{\raisebox{.14\height}{\scalebox{.7}{$\textstyle\sum$}}\cr\hidewidth$\textstyle\int$\hidewidth\cr}}   
  {\ooalign{\raisebox{.2\height}{\scalebox{.6}{$\scriptstyle\sum$}}\cr$\scriptstyle\int$\cr}}
  {\ooalign{\raisebox{.2\height}{\scalebox{.6}{$\scriptstyle\sum$}}\cr$\scriptstyle\int$\cr}}  
}

\begin{document}

\title{Nuclear polarization corrections in the $\mu\,^4$He$^+$ Lamb
  shift}

\author{C.~Ji}
\email{jichen@triumf.ca}
\affiliation{TRIUMF, 4004 Wesbrook Mall, Vancouver, BC V6T 2A3, Canada}

\author{N.~Nevo Dinur}
\email{nir.nevo@mail.huji.ac.il}
\affiliation{Racah Institute of Physics, The Hebrew University, Jerusalem 91904, Israel}

\author{S.~Bacca}
\email{bacca@triumf.ca}
\affiliation{TRIUMF, 4004 Wesbrook Mall, Vancouver, BC V6T 2A3, Canada}
\author{N.~Barnea}
\email{nir@phys.huji.ac.il}
\affiliation{Racah Institute of Physics, The Hebrew University, Jerusalem 91904, Israel}

\date{\today}


\begin{abstract}
Stimulated by the proton radius conundrum, measurements of the 
Lamb shift
in various
light muonic atoms are planned at PSI. The aim is to extract the rms charge
radius with high precision, limited by the uncertainty in the nuclear polarization
corrections. 
We present an {\it ab-initio} calculation of the nuclear polarization for
$\mu\,^4$He$^+$  leading to an energy correction in the 2$S$-2$P$ transitions
of $\delta^{A}_{pol}=-2.47$ meV $\pm 6\%$. 
We use two different state-of-the-art 
nuclear Hamiltonians 
and utilize the Lorentz integral transform  with  
hyperspherical harmonics expansion as few-body methods.
We take into account the leading multipole contributions, plus 
Coulomb, relativistic and finite-nucleon-size corrections. Our main source
of uncertainty is the nuclear Hamiltonian,
which currently limits the attainable accuracy.
Our predictions considerably reduce the uncertainty with respect to  previous
 estimates and
 should be instrumental to the $\mu\,^4$He$^+$ experiment planned for 2013.
\end{abstract}


\maketitle

{\it Introduction} ---
Recent laser spectroscopy measurements of the muonic Hydrogen Lamb shift~\cite{Lamb_shift}
($2S$-$2P$ transition) at PSI have tremendously improved the
accuracy in determining the proton charge radius
$\bra r_p^2\ket^{1/2}$. Besides experimental precision, the accurate deduction of
$\bra r_p^2\ket^{1/2}$ heavily relies on theory. Theoretical estimates of
quantum electro-dynamics (QED), recoil, and nuclear structure corrections are
needed. 
The proton radius extracted at PSI~\cite{Pohl:2010zza,Antognini:1900ns} is 10
times more accurate than the value determined from electron Hydrogen, i.e.,
CODATA-2010~\cite{Mohr:2012tt}, but also deviates from it by $7\sigma$. This
discrepancy, coined the ``proton radius puzzle'', is challenging the
understanding of experimental systematic errors and of theoretical
calculations based on the standard model. Alternative explanations involving
physics beyond the standard model (e.g., lepton flavor universality
violations) have also been proposed (see \cite{Pohl_review} for a review). To
understand this puzzle, one possible strategy is to investigate atoms with
other nuclear charges $Z$ or mass numbers $A$,
and track the persistence or variation of this discrepancy~\cite{Antognini:2011zz}. 
Extending the Lamb shift measurements to other muonic atoms,
e.g., $\mu\,$D, $\mu\,^3$He$^+$ and $\mu\,^4$He$^+$,
must be complemented by corresponding theoretical calculations.
Lamb shifts in light muonic atoms are very sensitive to nuclear structure
effects since a muon is 206 times heavier than an electron
and thus interacts more closely with the nucleus~\cite{Borie:1982ax, Borie:2012zz}.
The 2$S$-2$P$ energy difference
can be generally related to the nuclear charge radius
$\bra R^2_c\ket^{1/2}$ (in $\hbar=c=1$ units) by~\cite{Friar:1979zz},   
\begin{equation} \label{eq:E2s2p}
\Delta E \equiv \delta_{QED}+\delta_{pol} + \delta_{Zem} 
                + m_r^3 (Z\alpha)^4 \bra R^2_c\ket/12, 
\end{equation}
in an expansion of $Z\alpha$ up to $5^{th}$ order. 
Here 
$\alpha$ is the fine-structure constant 
and $m_r=m_{\mu} M_A /(m_{\mu}+M_A)$ is the reduced mass
related to the nuclear mass $M_A$ and the muon mass $m_{\mu}$.
$\delta_{Zem}$ is the 3$^{rd}$
Zemach moment~\cite{Zemach:1956zz} defined via the nuclear charge density, $\rho_0(\bs{R})$, as 
\begin{equation}
\label{eq:zemach3}
\delta_{Zem} = - \frac{m_r^4}{24}(Z\alpha)^5 \iint d \bs{R} d \bs{R}' \left|\bs{R}-\bs{R}'\right|^3 \rho_0(\bs{R})\rho_0(\bs{R}').
\end{equation}
Contributions to $\delta_{QED}$ in Eq.~\eqref{eq:E2s2p} are from vacuum
polarization, muon self energy and relativistic recoil; while $\delta_{pol} = \delta^{A}_{pol}+\delta^{N}_{pol}$ is the sum of the nuclear polarization $\delta^{A}_{pol}$ and the intrinsic nucleon polarizability $\delta^{N}_{pol}$. 
Since calculations of $\delta_{QED}$
and spectroscopy measurements of $\Delta E$ have both achieved high
accuracy, the current bottleneck in accurately extracting
$\bra R^2_c\ket^{1/2}$
from Eq.~\eqref{eq:E2s2p} lies in the polarization uncertainty. 
In muonic helium, to determine the nuclear radii with a relative accuracy of $3 \times
10^{-4}$,  $\delta_{pol}$ needs to be known at the $\sim 5\%$ level~\cite{Antognini:2011zz}.
Here we focus on the nuclear polarization $\delta^{A}_{pol}$. $\delta^{N}_{pol}$ depends on the internal nucleon structure and can be evaluated separately~\cite{Carlson:2011zd, Birse:2012eb, Miller:2012ne} apart from nuclear dynamics.

\begin{figure}[htb]
\centerline{\resizebox*{4.4cm}{2cm}{\includegraphics[angle=0]{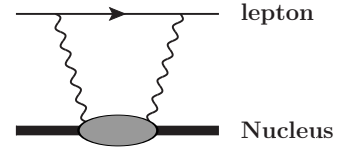}}}
\caption{
The lepton-nucleus two-photon exchange.
} 
\label{fig_diag} 
\end{figure}
The nuclear polarization is induced by a two-photon exchange process (Fig.~\ref{fig_diag}), 
where the nucleus in an atom is virtually excited by its
Coulomb interaction with the lepton.  
Effects  on the leptonic spectrum are evaluated in second-order perturbation
theory with inputs from  nuclear structure functions, also called
response functions. In early calculations
 structure functions were either calculated using simple nuclear
potentials  (e.g., $\mu\,$D~\cite{Lu:1993nq} and
$\mu$-$^{12}$C~\cite{Rosenfelder:1983aq}) or extracted from measurements of
photo-absorption cross sections (e.g., $\mu\,^4$He$^+$~\cite{Bernabeu:1973uf, Rinker:1976en, Friar:1977cf}). However, these approaches lack the desired accuracy. 
For example, Refs.~\cite{Bernabeu:1973uf, Rinker:1976en, Friar:1977cf}
           yielded \mbox{$\delta^{A}_{pol}=-3.1$ meV $\pm$ 20\%} for $\mu\,^4$He$^+$.
 Evaluations of the polarization effect
in $\mu\,$D  
using state-of-the-art potentials have significantly improved the
accuracy~\cite{Leidemann:1994vq, Pachucki:2011xr}. 
The purpose of this Letter is to extend these calculations to $\mu\,^4$He$^+$. 
We present the first {\it ab-initio} calculation of the nuclear polarization
effects in $\mu\,^4$He$^+$ using modern nuclear potentials. 
We consider systematically all  terms contributing to order
$(Z\alpha)^5$ and estimate the theoretical error. 

{\it Polarization Contributions} ---
Following 
works on $\mu\,$D by Pachucki~\cite{Pachucki:2011xr} and Friar~\cite{Friar:2013rha}
we separate contributions to the $\mu\,^4$He$^+$ polarization 
into non-relativistic, relativistic, Coulomb distortion and nucleon-size effects.   
The $\mu\,^4$He$^+$ system is described as a muon interacting with the $^4$He nucleus
containing four point-like nucleons by
\begin{equation}
\label{eq:H-muHe}
   H = H_{nucl} +H_\mu -\Delta H \,,
\end{equation}
where $H_{nucl}$ denotes the nuclear Hamiltonian,
and $H_\mu$ is the non-relativistic Hamiltonian
of a muon in the Coulomb potential of a point-like nucleus
\begin{equation}
\label{eq:H-muon-nrel}
   H_\mu = p^2/2m_r - Z\alpha/r\;.
\end{equation}
Here $p=|\bs p |$ ($r=|\bs r |$) is the relative momentum (distance) of the muon from
the center of mass (CM) of the nucleus. The last term in Eq.~\eqref{eq:H-muHe},  
\begin{equation}\label{eq:dH}
\Delta H = \sum_a^Z \Delta V(\bs{r},\bs{R}_a)\equiv 
         \sum_a^{Z} \alpha \left(\frac{1}{\left|\bs{r}-\bs{R}_a\right|} -\frac{1}{r}\right),
\end{equation}
represents the difference between the muon interaction with the nucleus and
the sum of Coulomb interactions between the muon and each
 proton, 
located  at a distance $\bs{R}_a$  from the CM.
Polarization effects are evaluated as corrections due to
$\Delta H$ in
second-order perturbation theory.
Utilizing the point-nucleon charge density operator  
\begin{equation}
\label{eq:rhoR-op}
\hat{\rho}(\bs{R}) \equiv \frac{1}{Z} \sum\limits_{a}^{Z} \delta(\bs{R}-\bs{R}_a),
\end{equation}
the nuclear polarization correction assumes the form
\begin{equation}\label{eq:dE-pol}
  \delta^{A}_{pol} = -\SumInt \limits_{N\neq N_0} \!\!\iint d\bs{R}d\bs{R}'
     \rho_N^*(\bs{R})P(\bs{R},\bs{R}',\omega_N) \rho_N(\bs{R}')\,,
\end{equation}
where
$\rho_N(\bs{R}) = \bra N | \hat{\rho}(\bs{R}) |N_0\ket$ 
is the charge density transition matrix element and
\begin{align}\label{eq:P-NR}
  P(\bs{R},\,&\bs{R}', \omega_N) = -Z^2 \int d\bs{r} d\bs{r}'
     \Delta V(\bs{r},\bs{R}) \bra \mu_0|\bs{r}\ket \cr &
     \bra \bs{r}
       |\frac{1}{H_\mu +\omega_N-\epsilon_{\mu_0}} | \bs{r}'\ket \bra \bs{r}'
       |\mu_0\ket \Delta V(\bs{r}',\bs{R}')
\end{align}
is the muonic matrix element.
Here $\omega_N=E_N-E_{N_0}$, and $E_{N_0}$, $E_{N}$, $|N_0\ket$ and $|N\ket$
are the nuclear ground- and excited-state energies and wave-functions
, respectively.
The symbol $\SumInt$ indicates a sum over discrete 
 plus an integration 
 over
continuum states.
$\epsilon_{\mu_0}$ and $|\mu_0\ket$ are 
the unperturbed atomic energy and wave-function in either the 2$S$ or 2$P$ state.
In Eq.~\eqref{eq:dE-pol} the nucleus is excited into all possible
intermediate states, which represents the inelastic part of the two-photon
exchange; while the elastic part is known as a finite-size
effect~\cite{Friar:1979zz}. 

The leading  contribution to $\delta^{A}_{pol}$ is
obtained  in the non-relativistic limit,  
neglecting in Eq.~\eqref{eq:P-NR} the Coulomb-potential part of $H_\mu$.
Only contributions to the 2$S$ state are considered, as 2$P$-state effects enter
only at order $(Z\alpha)^6$.
In this limit, we have 
\begin{align}
\label{eq:PNR}
 P(\bs{R},\bs{R}',\omega)=&
-Z^2 \phi^2(0) \int \frac{d\bs q}{(2\pi)^3}
\left(\frac{4\pi\alpha}{q^2}\right)^2 \left(1-e^{i\bs q\cdot \bs R}\right) 
\nn
&\times \frac{1}{q^2/2 m_r +\omega} \left(1-e^{-i\bs q\cdot \bs{R}'}\right),
\end{align}
where $\phi^2(0)=(m_rZ\alpha)^3 /8\pi$ is the normalization
coefficient of the muon $2S$ state. 
After integrating  over $\bs q$
in Eq.~\eqref{eq:PNR},
 terms not depending on both $\bs{R}$ and $\bs{R}'$ drop out,
due to the orthogonality of the nuclear eigenstates.
The resulting muonic matrix element $P$
is then a function of $\xi\sqrt{2m_r \omega}$,
  with $\xi\equiv|\bs{R}-\bs{R}'|$.
  Expanding $P$ in powers of $\xi\sqrt{2m_r \omega}$ up to 4$^{th}$ order yields
\begin{equation}
\label{eq:PNR-exp}
P(\xi,\omega) \simeq   \frac{m_r^3 (Z\alpha)^5}{12} 
\sqrt{\frac{2m_r}{\omega}} 
\left[\xi^2 - \frac{\sqrt{2 m_r \omega}}{4} \xi^3
+ \frac{m_r \omega}{10} \xi^4 \right].
\end{equation}
  $\xi$ indicates
  the ``virtual'' distance a proton travels
  during the two-photon exchange.
  According to the uncertainty principle
  it is related to $\omega$ by
  \mbox{$\xi \sim 1/\sqrt{2M_A \omega}$.} 
  Therefore the expansion parameter $\xi\sqrt{2m_r \omega}$ in Eq.~\eqref{eq:PNR-exp}
  is of order $\sqrt{m_r/M_A}\approx 0.17$.

In the following we will relate the different $\delta^{A}_{pol}$
terms coming from Eq.~\eqref{eq:PNR-exp} to structure functions. Details will
be given in a forthcoming paper~\cite{long_paper}. 
The structure functions are defined as
\begin{equation}
S_{O}(\omega) \equiv  \frac{1}{2J_0+1} \!\!\!\!
 \SumInt \limits_{N\neq N_0, J} \!\!\!\!
 |\bra N_0 J_0 ||\hat O ||N J\ket|^2
\delta(\omega-\omega_N),
\end{equation}
where $\hat O$ is a general operator.
Here we use the reduced matrix elements by employing the Wigner-Eckart theorem~\cite{Edmonds:1996}.
$J_0$ ($J$) is the total angular momentum of the ground (excited) state of $^4$He.

The leading contribution to the nuclear polarization,
       denoted by the superscript $(0)$,
       is the electric-dipole correction,
       which originates from the $\xi^2$ term in Eq.~\eqref{eq:PNR-exp}
\begin{equation}
\label{eq:del-R2}
\delta^{(0)}_{D1}  =-\frac{2\pi m_r^3}{9}  (Z\alpha)^5 \int^\infty_{\omega_{\rm th}}
d\omega \sqrt{\frac{2m_r}{\omega}} S_{D_1}(\omega), 
\end{equation}
where $\hat D_1 = \frac{1}{Z} \sum_a^Z R_a Y_1(\hat R_a)$, $Y_1$ is the
rank-1 spherical harmonics, and $\omega_{\rm th}$ 
indicates the threshold excitation energy of $^4$He, i.e., $\omega_{\rm
  th}=19.8$ MeV. 

The sub-leading
$\xi^3$ term is independent of $\omega$. Replacing 
$\SumInt_{N\neq N_0} |N\ket \bra N|$ with $1-|N_0\ket \bra N_0|$,
the contribution of this term, denoted by the superscript $(1)$, is 
\begin{align}
\label{eq:del-R3}
\delta^{(1)}=
   -&\frac{m_r^4}{24}(Z\alpha)^5\iint d \bs{R} d \bs{R}' |\bs{R}-\bs{R}'|^3 
\nn
&\times\left[\bra N_0| \hat{\rho}^\dagger(\bs{R})\hat{\rho}(\bs{R}')| N_0\ket -\rho_0(\bs{R})\rho_0(\bs{R}')\right],
\end{align}
where $\rho_0(\bs{R})\equiv \rho_{N_0}(\bs{R}) = \bra N_0 | \hat{\rho}(\bs{R})| N_0\ket$ is the
charge density, satisfying
$\int d\bs R \,\rho_0(\bs R)=1$.
It is convenient to write Eq.~\eqref{eq:del-R3} as
$\delta^{(1)}=\delta^{(1)}_{R3pp}+\delta^{(1)}_{Z3}$. 
The first term $\delta^{(1)}_{R3pp}$
is the ground-state expectation value of the proton-proton distance cubed.
The second term $\delta^{(1)}_{Z3}$
cancels exactly the 3$^{rd}$ Zemach moment $\delta_{Zem}$
that appears
in the finite-size corrections to the Lamb shift~\eqref{eq:E2s2p}.
This cancellation was also found by Pachucki~\cite{Pachucki:2011xr} and
Friar~\cite{Friar:1997tr, Friar:2013rha} in $\mu$D. 
Here we retain this term 
and calculate $\delta^{(1)}_{R3pp}$ and $\delta^{(1)}_{Z3}$ first in the
point-nucleon limit and then add finite-nucleon-size corrections.

Contributions from the sub-sub-leading $\xi^4$
term, denoted with the superscript $(2)$, are
\begin{align}
\label{eq:del-R4}
\delta^{(2)} &= \frac{m_r^5}{18} (Z \alpha)^5  \int^\infty_{\omega_{\rm th}}
d\omega \sqrt{\frac{\omega}{2m_r}} 
\nn
&\times\left[ S_{R^2}(\omega)
+ \frac{16\pi}{25} S_{Q}(\omega)
+ \frac{16\pi}{5}  \mathcal{S}_{D_1 D_3}(\omega) \right],
\end{align}
  where $S_{R^2}$ and $S_{Q}$ are the respective
  structure functions of the
  monopole $\hat R^2 = \frac{1}{Z} \sum_a^Z R^2_a$ and quadrupole
  \mbox{$\hat Q = \frac{1}{Z} \sum_a^Z R^2_a  Y_2(\hat R_a)$} operators. 
$\mathcal{S}_{D_1D_3}$ indicates the interference between two
multipolarity-1 operators $\hat D_1$ and $\hat D_3 =\frac{1}{Z} \sum_a^Z R^3_a
Y_1(\hat R_a)$ and is calculated as
\begin{equation}
\mathcal{S}_{D_1 D_3}(\omega) = \frac{1}{2} \left[S_{D_1+D_3}(\omega) -
  S_{D_1}(\omega) - S_{D_3}(\omega)\right]. 
\end{equation}
Effects from $\hat R^2$, $Q$ and the interference term in
Eq.~\eqref{eq:del-R4} are defined respectively as $\delta^{(2)}_{R2}$, $\delta^{(2)}_{Q}$
and $\delta^{(2)}_{D1D3}$. 

Since the electric-dipole contribution $\delta^{(0)}_{D1}$ dominates in the
 non-relativistic approximation, we add 
 relativistic corrections solely to this term. These corrections can
 be obtained from the longitudinal (L) and transverse (T) parts of the two-photon exchange
 amplitude~\cite{Bernabeu:1973uf, Rosenfelder:1983aq}.
 Replacing the non-relativistic Green's function with relativistic expressions, 
 we obtain relativistic corrections as
\begin{align}
\label{eq:ELT-full}
\delta^{(0)}_{L(T)}
=&\frac{2 m_r^3}{9} (Z\alpha)^5 \int^\infty_{\omega_{\rm th}} d\omega\,
\Theta_{ L(T)} \left(\frac{\omega}{m_r}\right)S_{D_1}(\omega), 
\end{align}
where the two energy-dependent weights are
\begin{align}
\label{eq:theta-L}
\Theta_L(\lambda) =& \pi \sqrt{2/\lambda} +2 \mathcal{F}(\lambda),
\\
\label{eq:theta-T}
\Theta_T(\lambda)=& \lambda+ \lambda \ln \left(2\lambda\right) + \lambda^2 \mathcal{F}(\lambda),
\end{align}
with
\begin{align}
\mathcal{F}(\lambda) =&
\sqrt{(\lambda-2)/\lambda}\; \rm{arctanh} \left(\sqrt{(\lambda-2)/\lambda}\right) 
\nn
&- \sqrt{(\lambda+2)/\lambda}\; \rm{arctanh} \left(\sqrt{\lambda/(\lambda+2)}\right), 
\end{align}
and $\lambda\equiv\omega/m_r$ ranges from $\sim 0.2$ to infinity. 
 Expressions similar to Eqs.~\eqref{eq:theta-L} and \eqref{eq:theta-T}
 are also derived by Martorell~{\it et al.}~\cite{Martorell:1995zz}, 
 whose transverse form is, however, valid only for $\lambda \geq 2$.

By including a Coulomb distorted muonic wave-function in the intermediate
state of the two-photon exchange in Fig.~\ref{fig_diag}, 
$\delta^{(0)}_{D1}$ is corrected in both the 2$S$ and 2$P$ states. We follow the derivation by
Friar~\cite{Friar:1977cf} and provide Coulomb-distortion corrections up 
to 2$^{nd}$ order in a
$Z\alpha\sqrt{2 m_r/\omega}$ expansion.
The Coulomb-distortion correction is given as the difference between the $2S$ and $2P$ levels:
\begin{align}
\label{eq:delta_C_0}
\delta_{C}^{(0)} =& -\frac{2\pi m_r^3}{9}  (Z\alpha)^6
             \int^\infty_{\omega_{\rm th}} d\omega 
\left[\frac{m_r}{\omega} \left(\frac{1}{6}+\ln\frac{2m_r Z^2
    \alpha^2}{\omega}\right)\right. 
\nn
& \left.-\frac{17}{16}Z\alpha \left(\frac{2m_r}{\omega}\right)^{3/2}\right]
S_{D_1}(\omega).
\end{align}
Even though $\delta_{C}^{(0)}$ is of order $(Z\alpha)^6$, its contribution is significantly enhanced by the $2S$-logarithmic term in Eq.~\eqref{eq:delta_C_0}.

Considering the finite size of the nucleons, the proton position in 
Eq.~\eqref{eq:dH} should be replaced by a convolution over the proton 
charge density, and a similar term should be added for the neutron.
For the proton and neutron form factors~\cite{Friar:2005je} 
we use low-momentum approximations:
$G^E_p(q^2)\simeq1-2q^2/\beta^2$ and $G^E_n(q^2)\simeq\lambda q^2$. 
Following Ref.~\cite{Friar:2013rha}, we choose 
$\beta=4.120\;\rm{fm}^{-1}$ and 
$\lambda=0.01935\;\rm{fm}^2$ which reproduce 
$\bra r_p^2 \ket^{1/2} = 0.8409$ fm~\cite{Antognini:1900ns} and 
$\bra r_n^2 \ket = -0.1161$~fm$^2$ \cite{Beringer:1900zz}. 
Since corrections to $\delta^{(0)}$ vanish, the leading nucleon-size (NS) correction enters in $\delta^{(1)}$ as 
\begin{align}\label{eq:delta_NS}
  \delta^{(1)}_{NS}
   =& - m_r^4(Z \alpha )^5 \left[\frac{2}{\beta^2}-\lambda \right] 
       \iint d\bs{R}d\bs{R}' |\bs{R}-\bs{R}'|
       \nn
       &\times
       \left[\bra N_0| \hat{\rho}^\dagger(\bs{R})\hat{\rho}(\bs{R}')| N_0\ket
         -\rho_0(\bs{R})\rho_0(\bs{R}')\right]\,,
\end{align}
where we have used the isospin symmetry of the $^4$He ground-state~\cite{Pisa05,Pisa08}.
The prefactors $2/\beta^2$ and $- \lambda$
account for respective contributions from
proton-proton and neutron-proton correlations, 
whereas neutron-neutron correlations are neglected.
Similarily to $\delta^{(1)}$, contributions to $\delta^{(1)}_{NS}$ from 
 the two integrands in Eq.~\eqref{eq:delta_NS} 
 are denoted as
$\delta^{(1)}_{NS}=\delta^{(1)}_{R1pp}+\delta^{(1)}_{Z1}$.

The sub-leading nucleon-size correction enters in $\delta^{(2)}$ as
\begin{equation}
\label{eq:delta_D1_2}
\delta^{(2)}_{NS} = - \frac{16\pi}{9} m_r^5 (Z \alpha )^5 \left[\frac{2}{\beta^2}-\lambda \right] 
       \int^\infty_{\omega_{\rm th}}
d\omega \sqrt{\frac{\omega}{2m_r}} S_{D_1}(\omega).
\end{equation}

Summing up the nuclear polarization corrections to the Lamb shift we have
\begin{align}\label{eq:pol-sum}
\delta^{A}_{pol} =&  \delta^{(0)}+\delta^{(1)}+\delta^{(2)} + \delta_{NS}
\nn
             =& \left[\delta^{(0)}_{D1} + \delta^{(0)}_{L} + 
                      \delta^{(0)}_{ T} + \delta^{(0)}_{ C}\right]
             +  \left[\delta^{(1)}_{R3pp} + \delta^{(1)}_{Z3}\right]
\nn
             +& \left[\delta^{(2)}_{R2} + \delta^{(2)}_{Q2} + \delta^{(2)}_{D1D3}\right]
             +  \left[\delta^{(1)}_{R1pp} + \delta^{(1)}_{Z1} + \delta^{(2)}_{NS} \right] 
. 
\end{align}

{\it Computational Tools} ---
The $^4$He structure functions involve a sum over all the 
spectrum, including energies beyond the 
three-body disintegration threshold. Thus, we calculate them
using the Lorentz integral transform  (LIT) method~\cite{EFROS94,REPORT07}, 
which allows exact calculations in this energy range. We use the effective
interaction hyperspherical harmonics (EIHH)~\cite{BaL00} 
few-body technique to solve the $^4$He ground state and the LIT equations. The
same methods were used, e.g., for the first realistic calculation of the $^4$He
dipole structure function in Ref.~\cite{Doron2006}.

For the nuclear Hamiltonian we use two state-of-the-art potential models that include three-nucleon (3N) forces:
{\it (i)}
the Argonne $v_{18}$~\cite{Wiringa:1994wb} nucleon-nucleon (NN) force
supplemented by the Urbana IX~\cite{PuP95} 3N force, denoted by AV18/UIX, and
{\it {(ii)}}
a  chiral effective field theory potential~\cite{Epelbaum:2008ga, 
Machleidt:2011zz}, denoted by $\chi$EFT, where the NN and 3N forces are at
N$^3$LO and N$^2$LO in the chiral expansion, respectively.
For the chiral 3N force we use the parameterization of the low-energy constants obtained
in~\cite{Na07} (\mbox{$c_D=1$} and \mbox{$c_E=-0.029$}). 
The calculated  $^4$He
binding energy, point-proton radius and electric-dipole polarizability $\alpha_E$ 
are respectively $28.422$ MeV, 1.432 fm and 0.0651 fm$^3$ for the AV18/UIX potential.
The corresponding numbers for the $\chi$EFT force are $28.343$ MeV, 1.475 fm and 0.0694 fm$^3$.
These numbers are in good agreement with previous calculations~\cite{Pisa08, tetrahedron, Stetcu09}.
The theoretical AV18/UIX ($\chi$EFT) binding energy and radius are respectively within $0.3\%$ ($0.1\%$)
and $3\%$ ($0.3\%$) of the experimental values. The uncertainty of $\alpha_E$, spanned by these two potentials, agrees with one in a recent study from variations of the $\chi$EFT low-energy constants~\cite{Stetcu09}, and is much smaller than the experimental error.

{\it Results} ---
We first check the formalism by comparing our $\mu$D results with
Pachucki~\cite{Pachucki:2011xr}. In Table~\ref{table:dipole-muD}, we present
all corrections related to the
dipole structure function $S_{D_1}(\omega)$ obtained from the AV18
potential~\cite{Winfried}. 
We find a good agreement for $\delta^{(0)}_{D1}$ and $\delta^{(0)}_{C}$. A difference in the relativistic corrections appears
because in Ref.~\cite{Pachucki:2011xr}  $\delta^{(0)}_L$  includes only the leading term of $\Theta_L$~\eqref{eq:theta-L} in an $\omega/m_r$
 expansion and neglects $\Theta_T$~\eqref{eq:theta-T}, since it
 is one-order higher in $\omega/m_r$.
 These higher-order terms, which we include, provide additional relativistic corrections to the
 Lamb shift in 
 $\mu D$. 
 Consequently, the $-1.680$ meV result of Ref.~\cite{Pachucki:2011xr} changes to $-1.698$ meV, 
where in this case the cancellation of the  3$^{rd}$ Zemach moment is
implemented as in 
Ref.~\cite{Pachucki:2011xr}. 
 
\begin{table}[htb]
\begin{center}
\caption{Nuclear polarization contributions to the $2S$-$2P$ Lamb shift
  $\Delta E$ [meV] in $\mu$D,
  compared to Pachucki~\cite{Pachucki:2011xr}. } 
\label{table:dipole-muD}
\begin{tabular}{l r r}
\hline
                     & Ref.~\cite{Pachucki:2011xr}& This work  \\
\hline
$\delta^{(0)}_{D1}$   & -1.910$~\;~$     & -1.907$~\;~$   \\
$\delta^{(0)}_{L}$    & 0.035$~\;~$      &  0.029$~\;~$   \\
$\delta^{(0)}_{T}$    &    -- $~\;~\;\;$     & -0.012$~\;~$   \\
$\delta^{(0)}_{C}$    &  0.261$~\;~$     & 0.259$~\;~$    \\
\hline
\end{tabular}
\end{center}
\end{table} 

Now we turn to $\mu\,^4$He$^+$ and discuss the first {\it ab-initio}
calculations for $\delta^{A}_{pol}$. 
Numerical results for the AV18/UIX and $\chi$EFT potentials
are presented in Table.~\ref{table:polar},
leading to an average value of 
$\delta^{A}_{pol}=-2.475$~meV.
In the point-nucleon treatment, we observe that the leading contribution 
 $\delta^{(0)}$, amounting to $-3.743$~meV with AV18/UIX 
and $-3.981$~meV with $\chi$EFT, 
strongly dominates in $\delta^{A}_{pol}$. 
 \begin{table}[htb]
 \begin{center}
 \caption{Nuclear polarization contributions to the $2S$-$2P$ Lamb
          shift $\Delta E $ [meV] in $\mu\,^4$He$^+$.} 
 \label{table:polar}
 \begin{tabular}{l@{\hspace{6mm}} l@{\hspace{8mm}} r r}
 \hline
 &                          & AV18/UIX  & $\chi$EFT$\;$ \\
 \hline
 \multirow{4}{*}{$\delta^{(0)}$}
 &$\delta^{(0)}_{D1}$       & -4.418$~\;~\;$                & -4.701$\;$    \\
 &$\delta^{(0)}_{L}$    &  0.289$~\;~\;$                &  0.308$\;$    \\
 &$\delta^{(0)}_{T}$    & -0.126$~\;~\;$                & -0.134$\;$    \\
 &$\delta^{(0)}_{C}$    &  0.512$~\;~\;$                &  0.546$\;$    \\
 \hline
 \multirow{2}{*}{$\delta^{(1)}$}
 &$\delta^{(1)}_{R3pp}$     & -3.442$~\;~\;$                & -3.717$\;$     \\
 &$\delta^{(1)}_{Z3}$       &  4.183$~\;~\;$                &  4.526$\;$     \\
 \hline
 \multirow{3}{*}{$\delta^{(2)}$}
 &$\delta^{(2)}_{R2}$       &  0.259$~\;~\;$                &  0.324$\;$     \\
 &$\delta^{(2)}_{Q}$        &  0.484$~\;~\;$                &  0.561$\;$     \\
 &$\delta^{(2)}_{D1D3}$     & -0.666$~\;~\;$                & -0.784$\;$     \\
 \hline
 \multirow{3}{*}{$\delta_{NS}$}
 &$\delta^{(1)}_{R1pp}$     & -1.036$~\;~\;$                & -1.071$\;$    \\ 
 &$\delta^{(1)}_{Z1}$       &  1.753$~\;~\;$                &  1.811$\;$    \\ 
&$\delta^{(2)}_{NS}$        & -0.200$~\;~\;$                & -0.210$\;$    \\ 
 \hline
 \multicolumn{2}{l}{$\delta^{A}_{pol}$}
                            & -2.408$~\;~\;$                & -2.542$\;$   \\
 \hline
 \end{tabular}
 \end{center}
 \end{table}

Regarding the sub-leading terms, each individual term in
  $\delta^{(1)}$ (or $\delta^{(2)}$) is not necessarily small. 
  Only their complete combination at each order fulfills the expansion 
  in $\xi\sqrt{2m_r\omega} \sim\!\! \sqrt{m_r/M_A}$
  as a consequence of the uncertainty principle, and yields
  \mbox{$\delta^{(1)} = 0.775$~meV} and \mbox{$\delta^{(2)} = 0.089$~meV} when averaging AV18/UIX and $\chi$EFT calculations. 
  As expected, $\delta^{(1)}$ and $\delta^{(2)}$ are respectively one- and two-order smaller in $\sqrt{m_r/M_A}$ than $\delta^{(0)}$.
The nucleon-size correction contributes an additional $\delta_{NS}=0.523$~meV
in  average. The latter depends on the value of $\bra r_p^2\ket^{1/2}$: using 0.8775 fm~\cite{Mohr:2012tt} will increase $\delta_{NS}$ to $0.579$ meV.

The numerical accuracy of $\delta^{A}_{pol}$ is also studied.
 The error in the EIHH method is controlled by the convergence with respect to the maximum
 grand-angular momentum, $K_{max}$, which determines the size of the 
 model space~\cite{BaL00}.  
 This error, obtained by taking the difference between results
 with
 $K_{max}=22~(20)$ to those with $K_{max}-4$, is $0.4\%$ ($0.2\%$) 
 for AV18/UIX ($\chi$EFT).
 An additional $0.2\%$ error is estimated
 by comparing the results from integrating the structure functions 
 calculated using the LIT method, with those obtained by a Lanczos sum-rule method as in
 Ref.~\cite{tetrahedron}. 

The difference in $\delta^{A}_{pol}$ that comes from using the AV18/UIX or $\chi$EFT
potential amounts to 0.134 meV
and represents the uncertainty in
nuclear physics. 
 Both potentials are tuned to fit the $^3$H binding energy,
  and they reproduce the $^4$He binding energy to few parts per mil.
  They differ, however, in their respective predictions for the nuclear charge radius.
  Given the relations between the structure functions and the charge radius \cite{tetrahedron},
  it is plausible that the uncertainty in $\delta^{A}_{pol}$ can be reduced
  using the $^4$He charge radius to constrain the nuclear potential models.
This systematic uncertainty dominates the
errors in predicting $\mu\,^4$He$^+$ 
polarization effects.  
The difference between the two models divided by $\sqrt{2}$ gives a
$4\%$ error, which can be interpreted  
as a $1\sigma$ deviation from the central value. 
The magnetic polarization is negligible in $^4$He~\cite{ee'MT13}.
Terms of order $(Z\alpha)^6$, relativistic corrections to polarizations other
than dipole, and higher-order nucleon-size effects will be explored in the future. 
The sum of all these additional corrections is expected to be a few percent. 
In a quadratic sum of all the errors mentioned above we estimate the accuracy of our calculation to be $\pm 6\%$.
We did not include the contribution from the disputed intrinsic nucleon 
polarizability~\cite{Carlson:2011zd, Birse:2012eb, Miller:2012ne}, because it can be estimated
independently of the nuclear Hamiltonian (see e.g.~\cite{Pachucki:2011xr,Friar:2013rha}).

{\it Conclusions} ---
We perform the first {\it ab-initio} calculation for the  $\mu\,^4$He$^+$
polarization correction  obtaining 
$\delta^{A}_{pol}=-2.47$ meV $\pm 6\%$.    
This result significantly improves the accuracy and is close to the upper bound of
previous predictions 
$\delta^{A}_{pol}=-3.1$ meV $\pm 20\%$~\cite{Bernabeu:1973uf, Rinker:1976en, Friar:1977cf}.
The theoretical accuracy is limited by the uncertainty in the nuclear Hamiltonian,
which is probed by using two different state-of-the-art nuclear potentials.
 Exploring other choices for potential parameterizations
and including higher-order $\chi$EFT forces can possibly narrow this uncertainty.  
Our result allows a significant improvement in the precision
of $\bra R^{2}_c \ket$ that will be extracted from the $\mu\,^4$He$^+$ Lamb shift measurements planned for 2013.

\begin{acknowledgments}
  We thank Winfried Leidemann for providing us
  with the AV18 deuteron dipole structure function.
  We acknowledge constructive discussions with Jim Friar, Franz Kottmann, Randolf Pohl and Gerald A. Miller.
  This work was supported in parts by the Natural Sciences
  and Engineering Research Council (NSERC), the National Research
  Council of Canada, and the Israel Science Foundation (Grant number
  954/09).
\end{acknowledgments}


\end{document}